\documentclass[11pt,a4paper]{article}
\usepackage{jheppub}
\usepackage{etoolbox}

\def\one{{\,\hbox{1\kern-.8mm l}}}
\newcommand{\Dslash}{\not{\hbox{\kern-4pt $D$}}}
\newcommand{\pdslash}{\not{\hbox{\kern-2pt $\partial$}}}

\newcommand{\Comment}[1]{{}}

\def\IZ{{\mathbb Z}}

\def\IR{{\mathbb R}}

\newcommand{\bc}{\begin{center}}
\newcommand{\ec}{\end{center}}
\newcommand{\ba}{\begin{array}}
\newcommand{\ea}{\end{array}}
\newcommand{\beq}{\begin{equation}}
\newcommand{\eeq}{\end{equation}}
\newcommand{\bea}{\begin{eqnarray}}
\newcommand{\eea}{\end{eqnarray}}
\newcommand{\bmx}{\begin{pmatrix}}
\newcommand{\emx}{\end{pmatrix}}
\newcommand{\nn}{\nonumber}

\newcommand{\be}{\begin{equation}}
\newcommand{\ee}{\end{equation}}

\newcommand{\del}{\partial}
\newcommand{\half}{{\frac{1}{2}\,}}

\newcommand{\tD}{{\tilde D}}
\newcommand{\tdel}{{\tilde \del}}

\newcommand{\tN}{{\tilde N}}
\newcommand{\tll}{{\tilde \ell}}

\newcommand{\eref}[1]{Eq.\,(\ref{#1})}

\newcommand{\taubar}{{\bar \tau}}

\newcommand{\tchi}{{\tilde \chi}}

\newcommand{\dG}{{\,{\rm dim}~G}}
\def\IB{\relax{\rm I\kern-.18em B}}
\def\IC{{\relax\hbox{\kern.3em{\cmss I}$\kern-.4em{\rm C}$}}}
\def\ID{\relax{\rm I\kern-.18em D}}
\def\IE{\relax{\rm I\kern-.18em E}}
\def\IF{\relax{\rm I\kern-.18em F}}
\def\II{\relax{\rm I\kern-.18em I}}
\def\IZ{\relax{\sf Z\kern-.35em Z}}
\def\Id{\relax{1\kern-.32em 1}}
\def\IG{\relax\hbox{$\inbar\kern-.3em{\rm G}$}}
\def\IR{\relax{\rm I\kern-.18em R}}
\newcommand\sfrac[2]{{\textstyle\frac{#1}{#2}}}

\newcommand\shalf{{\textstyle\frac12}}

\title{On 2d Conformal Field Theories with Two Characters}

\author{Harsha R. Hampapura}
\author{and Sunil Mukhi}

\affiliation{Indian Institute of Science Education and Research,\\
Homi Bhabha Rd, Pashan, Pune 411 008, India}

\emailAdd{harshahr93@gmail.com}
\emailAdd{sunil.mukhi@gmail.com}

\abstract{Rational CFT's are classified by an integer $\ell$, the number of zeroes of the Wronskian of their characters in moduli space. For $\ell=0$ they satisfy non-singular modular-invariant differential equations, while for $\ell>0$ the corresponding equations have singularities. We survey CFT's with two characters and $\ell=0,2,3,4$ and verify the consistency, at the level of characters, of some candidate theories with $\ell\ne 0$. For $\ell=2$ there are seven consistents sets of characters. We identify specific combinations of level-1 current algebras that are potential symmetries of the corresponding CFT's.}

\preprint{}


\makeatletter
   \patchcmd{\maketitle}{\@fpheader}{\includegraphics[height=15mm]{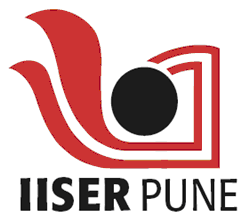}}{}{}
  \makeatother
    
\begin{document}

\maketitle

\section{Introduction}

The characters of rational conformal field theories in 2 dimensions (see for example Ref.\cite{Anderson:1987ge}) are a set of $p$ holomorphic functions $\chi_i(\tau)$ on the moduli space of the torus, such that the CFT partition function can be written as a bilinear:
\be
Z(\tau,\taubar)=\sum_{i,j=0}^{p-1} M_{ij}\chi_i(\tau)\chi_j(\taubar)
\ee
 The characters have a power series expansion:
\be
\chi_i(\tau)=\sum_{n=0}^\infty a_n^{(i)} q^{\alpha_i+n},\quad i=0,1,\cdots,n-1
\ee
where $q=e^{2\pi i\tau}$. Here $\alpha_i$ are the exponents, corresponding to $h_i-\frac{c}{24}$ where $h_i$ are the conformal dimensions of the primaries and $c$ is the central charge. The index $i=0$ corresponds to the identity character, for which $h_0=0$, while the remaining indices label characters over different primaries of the theory. If there are special symmetries such as complex conjugation or triality then a single character can correspond to multiple primaries (background material on these points and the following material can be found in Refs.\cite{Mathur:1988rx,Mathur:1988na,Mathur:1988gt, Naculich:1988xv} and references therein).

The partition function is required to be modular invariant:
\be
Z(\gamma \tau,\gamma\taubar) = Z(\tau,\taubar),\quad \gamma=
\begin{pmatrix}
~a~ & ~b~\\ ~c~ & ~d~
\end{pmatrix}
\in SL(2,Z)
\ee
This is ensured if the characters transform as vector-valued modular forms:
\be
\chi_i(\gamma\tau)= \sum_kV_{ik}(\gamma)\chi_k(\tau)
\label{vvmf}
\ee
where $V_{ik}$ are unitary matrices. In this case it is evident that the diagonal combination with $M_{ij}=\delta_{ij}$ always furnishes a modular invariant. If one character corresponds to more than one primary then this degeneracy factor must be included, $V_{ik}$ is then not unitary but can be enlarged to a bigger unitary matrix. Depending on the theory, other invariants than the diagonal one may exist (for details, see \cite{DiFrancesco:1997nk}).

The problem of classifying all possible theories with a single character ($p=1$) has been addressed as follows. In this case the partition function is the modulus-squared of the identity character $\chi_0(\tau)$. In turn, this character is completely determined by the Klein $j$-function:
\be
j(\tau) =\frac{(E_4)^3}{\Delta},\quad \Delta=\frac{E_4^3-E_6^2}{1728}
\label{jdeltadef}
\ee
where $E_4, E_6$ are the holomorphic Eisenstein series defined by:
\be
E_{2k}(\tau)=\frac{1}{2\zeta(2k)}\sum_{\genfrac{}{}{0pt}{}{m,n\in \IZ}{(m,n)\ne (0,0)}}\frac{1}{(m\tau+n)^{2k}}
\label{eisen}
\ee
The allowed combinations of the $j$ function must be such that the resulting characters are single-valued, modular-invariant upto a possible phase, and have only non-negative-integer degeneracies of states above a non-degenerate vacuum state. All these requirements are satisfied by any function of the form\footnote{The factor of $(j-1728)^\half$ by itself does not satisfy positivity, so this case is excluded.}:
\be
\chi(\tau)=j^\delta (j-1728)^\beta P(j)
\label{genone}
\ee
where $\delta=0,\frac13,\frac23$, $\beta=0,\half$ and $P(j)$ is a polynomial in $j$ whose coefficients must be adjusted so that the $q$-expansion has non-negative coefficients.

It is not guaranteed that all such functions correspond to genuine conformal field theories. Some examples that are known to correspond to CFT's are the monomial $\chi=j^{\frac{n}{3}}$ which can be identified with the character of tensor-product CFT's of the form $(\hbox{Spin}(32)/Z_2)^a \otimes (E_8)^b$ with $2a+b=n$. Another combination that has been successfully identified is $j-744$, the character of the Monster CFT. It is less clear whether the remaining combinations describe genuine CFT's. For example Ref.\cite{Schellekens:1992db} lists 71 possible CFT's with $c=24$, although there are infinitely many potential characters with this  central charge, of the form $j+N, -744\le N<\infty$\footnote{We understand from the referee that a CFT has been rigorously constructed for only 69 of these cases. Also there may conceivably be multiple CFT's for a given chiral algebra, and even for the character without a chiral algebra. We thank the referee for these observations.}. Another example is that of polynomials in $j$ that are tuned to make certain coefficients vanish. These have been proposed to correspond to ``extremal'' CFT's \cite{Witten:2007kt}. However, using mathematical properties of CFT's \cite{Zhu:1996} and some genericity assumptions, some doubt has been cast \cite{Gaberdiel:2007ve,Gaberdiel:2008pr} on the actual existence of CFT's corresponding to these ``extremal characters'', at least for sufficient large central charge. 

From this point of view, the starting point to discover a CFT is to find candidate characters with the desired modular properties. Motivated by the above observations about single-character theories, we would like to know the status of conformal field theories with two characters. In the present work we will look for pairs of characters satisfying the minimal necessary conditions for a CFT, namely that they transform into each other under modular transformations and have power-series expansions in $q=e^{2\pi i\tau}$ (where $\tau$ is the modular parameter of the torus) with non-negative integer coefficients (apart from an overall fractional power). Also, the identity character must have a non-degenerate ground state. We will use the technique of modular-invariant differential equations, which was employed in Ref.\cite{Mathur:1988na} and subsequently in Refs.\cite{Naculich:1988xv} and \cite{Kiritsis:1988kq}, to classify a sub-family of two-character theories. 

We will encounter some interesting examples with $\ell=2$ that have been previously listed in Ref.\cite{Naculich:1988xv}. We verify that these satisfy the consistency conditions for CFT characters to high orders in the $q$-expansion. We also rule out the existence of theories at $\ell=3$ and discuss some general features for $\ell=4,5$. Next we turn to a detailed analysis of the $\ell=2$ case, finding striking parallels in terms of central charges, conformal dimensions, modular transformations and fusion rules with the more familiar $\ell=0$ examples of Ref.\cite{Mathur:1988na} which were successfully identified with Kac-Moody algebras at level $k=1$. We similarly try to identify the Kac-Moody algebras for the $\ell=2$ theories and find exotic combinations of level-1 algebras that appear to describe them well in terms of central charge and conformal dimensions.

\section{Modular-invariant differential equations and the Wronskian}

The modular transformation properties of characters, \eref{vvmf}, indicate that they arise as the independent solutions of a degree-$p$ modular-invariant differential equation. Such an equation must be of the form \cite{Anderson:1987ge,Eguchi:1988wh,Mathur:1988na}:
\be
\left(D^p + \sum_{k=0}^{p-1} \phi_k(\tau) D^k\right)\chi=0
\label{modinveq}
\ee
where $D$ is a covariant derivative including a connection term. For the equation to be modular invariant, each term in the differential operator must transform homogeneously under modular transformations with the same weight. To achieve this, we must take:
\be
D \equiv \del -\frac{i\pi r}{6}E_2(\tau)
\label{covdev}
\ee
where $\del\equiv \frac{\del}{\del\tau}$, $r$ is the modular weight of the object on which it acts, and $E_2(\tau)$ is a special Eisenstein series defined as:
\be
E_{2}(\tau)=1+\frac{1}{2\zeta(2)}\sum_{m\ne 0}\sum_{n\in\IZ}\frac{1}{(m\tau+n)^{2}}
\label{eisen.2}
\ee
$E_2$ transforms inhomogeneously (as a connection) under modular transformations and provides us the desired covariant derivative. Note that $D$ augments the modular weight by 2, so an expression like $D^n\chi$ is shorthand for:
\be
\left(\del - \frac{i\pi(n-1)}{3}E_2\right)\left(\del-\frac{i\pi(n-2)}{3}E_2\right)\cdots \left(\del-\frac{i\pi}{3}E_2\right)\del\chi
\ee
For \eref{modinveq} to be modular invariant, the coefficient functions $\phi_k(\tau)$ must be modular covariant of weight $2(p-k)$. It is well-known that holomorphic modular forms are spanned by linear combinations of expressions of the form $(E_4)^a (E_6)^b$ for arbitrary non-negative integers $a$ and $b$, whose weight is $4a+6b$. However, as we will shortly see, we need not require that the $\phi_k$ are holomorphic. These coefficient functions can be meromorphic, in which case there are many more possibilities involving rational functions  of $E_4$ and $E_6$. 

To classify the possible cases, we note following Ref.\cite{Mathur:1988na} that the $\phi_k$ can be expressed as follows in terms of the independent solutions $\chi_0,\chi_1,\cdots \chi_{p-1}$ of the differential equation:
\be
\phi_k(\tau)=(-1)^{n-k}\frac{W_k}{W}
\ee
where $W_k$ are the Wronskian determinants:
\be
W_k\equiv 
\begin{Vmatrix}
\chi_0 & \chi_1 &\cdots &  \chi_{p-1}\\
D \chi_0 &D\chi_1 &\cdots &D \chi_{p-1}\\
\vdots & &&\vdots \\
D^{k-1}\chi_0& D^{k-1}\chi_1 & \cdots & D^{k-1}\chi_{p-1}\\
 D^{k+1}\chi_0 & D^{k+1}\chi_1 &\cdots & D^{k+1}\chi_{p-1}\\
 \vdots &&& \vdots\\
 D^{p}\chi_0 &D^{p}\chi_1 &\vdots &D^{p}\chi_{p-1}\\
\end{Vmatrix}
\ee
and $W=W_p$. When we refer to ``the Wronskian'' we will always mean $W$. Now, the characters themselves must be holomorphic everywhere in moduli space. Thus the only way that $\phi_k$ can be singular is for $W$ to have zeroes. Clearly the number of poles of $\phi_k$ is bounded above by the number of zeroes of $W$.

This leads us to classify differential equations by the number of zeroes of $W$. Because the torus moduli space has two orbifold points: $\tau=e^{\frac{i\pi}{3}}$ of order $\frac13$ and $i$ of order $\half$, the allowed numbers of zeroes of $W$ are $\frac{\ell}{6}$ with $\ell=0,2,3,4,\cdots$. This integer $\ell$ will play a key role in what follows. For a definite order $p$ of the differential equation and a definite $\ell$, we can completely determine the monomials in $E_4,E_6$ that are allowed in the equation. This reduces the unknowns to a finite number of real parameters. 

Having fixed $p$ and $\ell$, the strategy is to write down the most general differential equation consistent with it, insert a power series solution and determine the degeneracies of low-lying levels as functions of the independent parameters of the equation. As indicated earlier, the characters are required to have the power series expansion:
\be
\chi_i(\tau) = \sum_{n=0}^\infty a_n^{(i)} q^{\alpha_i+n}
\ee
There is an important relation \cite{Mathur:1988na} between the exponents $\alpha_i$, the order $p$ of the differential equation and the integer $\ell$ that we have just introduced:
\be
\sum_{i=0}^{p-1}\alpha_i = \frac{p(p-1)}{12}-\frac{\ell}{6}
\label{valenceform}
\ee
This relation arises from the Riemann-Roch theorem. As we will see, it determines one of the parameters in the differential equation. 

Since the $a_n^{(i)}$ are functions of the continuous parameters in the equation, they will not in general be integral or even positive. However for special values of the parameters they all come out to be non-negative integers. This is a very rare occurrence, and can rigorously be verified only by an infinite number of checks, namely computing $a_n^{(i)}$ for all $n$. In  practice we find that if for some values of the parameters the first few $a_n^{(i)}$ are non-negative then the remaining ones follow this trend to very high levels (say the first 5000 $a_n$'s). In this case we have a potential CFT. 

An additional complication comes from the fact that the differential equation only determines the ratios $\frac{a_n^{(i)}}{a_0^{(i)}}$. Here $a_0^{(i)}$ is the degeneracy of the ground state for a given character. For the leading (identity) character, corresponding to $i=0$, we must have $a_0^{(0)}=1$, corresponding to a non-degenerate vacuum. Therefore, for $i=0$ the calculated quantities  $\frac{a_n^{(i)}}{a_0^{(i)}}$ must come out to be integer. For the other (non-identity) characters with $i\ne 0$, it is sufficient for the ratio to be a rational number with a common denominator. The denominator can then be ascribed to the degeneracy $a_0^{(i)}$ of the primary state. Importantly the denominator should not keep growing with $n$ for a fixed $i$, because then there would be no way to fix the degeneracy of the ground state. Again, in practice we find that for a few very special values of the parameters in the equation, the non-identity degeneracies come out to be rational numbers with denominators built out of a small number of prime factors (typically 2 or 3 such factors) and this property remains stable to very high levels. Indeed in the favourable cases, once we have found certain denominators for the first few $a_n$ (say $n$ upto 5 or 6), then no new prime factors occur all the way up to $a_{5000}$. For comparison, in the unfavourable cases the denominators grow rapidly with the steady accumulation of new prime factors. Proceeding as above, we filter out ``most'' of the possible values of the parameters in the differential equation and are thereby restricted to a discrete set. These are our candidate CFT's. One can then apply more checks to the resulting characters to see if they correspond to a genuine CFT, for example, determining the fusion rules and chiral algebra.

In this paper we apply the above strategy to two-character theories with $\ell> 0$. To our knowledge, no two-character CFT with $\ell>0$ seems to have been explicitly constructed, other than tensor products involving a one-character theory and an $\ell=0$ theory (an example is discussed just below \eref{ruleout}). See also Ref.\cite{Milas:2004} for more discussion on theories with $\ell>0$. A set of candidate characters with $\ell=2$ were found in Ref.\cite{Naculich:1988xv} and we will reproduce them in our computations. However to our knowledge nothing more has been known about the underlying CFT's, if any. Note that the minimum $\ell$ for a tensor product theory is 4, because the minimum $\ell$ of a one-character theory is 2. Theories with more than two characters and $\ell>0$ are known, for example consider the orbifold line of the free boson at $c=1$ at a radius $\sqrt{2p}$. From the operator spectrum of these theories (see e.g. page 785 of \cite{DiFrancesco:1997nk}) it is easily verified that they have $p+4$ characters and $\ell=3(p-1)$. But this series starts with a minimum of five characters. 

There also exists a claim \cite{Kiritsis:1988kq} about the classification of {\em all} two-character theories with arbitrary $\ell$ in terms of a basic set of characters having $\ell=0$ (shown in Table I of Ref.\cite{Kiritsis:1988kq}), but some of which do not by themselves fulfil the requirements of CFT (specifically they have negative degeneracies)\footnote{We thank Elias Kiritsis for bringing this reference to our notice and for discussions.}. This result has been argued using theorems on monodromy representations and apparent singularities, and states that all sets of two characters are tensor products of any of these basic characters with a suitable function of $j(\tau)$. The product has to be chosen in such a way as to render all coefficients in the series expansion non-negative (which is not the case for all elements of the basic set). We notice that in particular, the work of Ref.\cite{Kiritsis:1988kq} seems to rule out the existence of two-character theories with $\ell=2$, which conflicts with the discovery of such theories both in Ref.\cite{Naculich:1988xv} and the present work.

\section{Tensor-product theories}

Before proceeding with the analysis, let us try to understand how the parameter $\ell$ defined above behaves when we take the tensor product of CFT's. This is important because, while discovering CFT's through differential equations, one always risks re-discovering the product of known theories. Since this issue has not been previously analysed as far as we know, we first consider some general situations and then specialise to the case of two characters. Consider a pair of theories, one with $p$ characters $\chi_i$, exponents $\alpha_i$ and some fixed $\ell$, and the other with $p'$ characters $\chi'_{i'}$, exponents $\alpha'_{i'}$ and some fixed $\ell'$. Generically, the tensor product is the CFT whose $pp'$ characters are $\tchi_{(ii')}=\chi_i\chi'_{i'}$ and the exponents are ${\tilde\alpha}_{(ii')}=\alpha_i+\alpha'_{i'}$. We would like to know the value of $\tilde\ell$ for this theory. Applying the relation \eref{valenceform}, we find:
\be
{\tilde\ell}= \half pp'(p-1)(p'-1)+p'\ell +p\ell'
\ee
This formula is not completely general, since we will see in a moment that it does not hold in the presence of degeneracies. But first let us assume two distinct, generic CFT's in which case the formula does hold. Then clearly $\ell$ increases upon tensoring the two theories. As an example suppose that the first theory is a one-character theory, $p=1$ with any $\ell$, while the second is any theory with $p'$ characters but $\ell'=0$. Then the above formula says that ${\tilde\ell}=p'\ell$. Since one-character theories always have $\ell> 0$, this simply tells us that the tensor product theory acquires a nonzero $\tilde\ell$. A more nontrivial class of examples arises when the two theories have $p,p'>1$ but $\ell=\ell'=0$. Then the product theory has non-vanishing $\tilde\ell = \half pp'(p-1)(p'-1)$. This appears to imply that if we study differential equations with $\ell=0$ and any number of characters, we will never encounter tensor products of theories with lower numbers of characters. However this is only true when we consider distinct theories in the product. The situation is different when we take tensor products of identical theories, because of the degeneracies mentioned above. Indeed, the above formula and its consequences are invalidated whenever two or more distinct primaries have degenerate conformal dimensions. One concrete class of examples is provided by tensoring any $p$-character CFT with itself. If the original theory has $p$ primaries then the product has $p^2$ primaries, but some of them are pairwise degenerate and there are only $\frac{p(p+1)}{2}$ independent characters. In this case, one finds a different formula:
\be
\tll= \frac{(p+1)p(p-1)(p-2)}{8} + (p+1)\ell
\label{neq.2}
\ee
If the original theory has more than $p$ distinct primaries of which some are degenerate, then the product theory has higher degeneracies but the  formula above continues to hold. 

In particular, the above discussion tells us that the square of a two-character theory with $\ell=0$ is a three-character theory that also has $\ell=0$. This phenomenon persists for higher tensor products. If we take the $n$th tensor product of a two-character CFT having a given value of $\ell$, then the resulting theory has $n+1$ characters and:
\be
{\tilde \ell}=\frac{n(n+1)}{2}\ell
\label{peq.2}
\ee
This immediately proves that the class of CFT's with $p$ characters and $\ell=0$ is non-empty for all $p$. It also tells us that when studying theories with $\ell=0$ and a fixed number of characters, we will definitely encounter powers of $\ell=0$ theories with lower numbers of characters.

Note also that if we take powers of theories with more than two characters then $\ell$ always increases. Consider the $n$th power of a $p$-character theory with some value of $\ell$. The resulting theory has $p^n$ primaries but only 
\be
{\tilde p}=\begin{pmatrix} p+n-1\\ p-1\end{pmatrix}
\label{tpeq}
\ee 
characters. To find the formula for ${\tilde \ell}$ of the composite theory, we need to know the sum over all exponents $\tilde\alpha_i$. This is found as follows. Consider all ordered partitions $\{n_i\}$ of $n$ into precisely $p$ numbers $n_i$ that take the possible values $0,1,2,\cdots,n$. Thus $n=\sum_{i=1}^p n_i$. Now let us pick the first integer $n_1$ and calculate $q=\sum_{\{n_i\}}n_1$. Clearly this number would be the same had we summed over any of the other $n_i$. $q$ counts how many times a particular exponent $\alpha_i$ of the original theory appears in the product theory. Therefore:
\be
\sum_j {\tilde \alpha}_j = q\sum_i \alpha_i
\ee
where $i$ runs over characters of the original theory and $j$ over characters of the product theory. It is straightforward to verify that 
\be
q=\begin{pmatrix}p+n-1\\ p\end{pmatrix}
\label{qeq}
\ee
Then the formula for ${\tilde\ell}$ of the composite theory is:
\be
\tll = \frac{{\tilde p}({\tilde p}-1)}{2}-q\,\frac{p(p-1)}{2} + q\ell
\ee
with $\tilde p$ and $q$ given by Eqs.(\ref{tpeq}) and (\ref{qeq}) respectively. 
One can easily check that this reproduces the formula in \eref{neq.2} for $n=2$ and arbitrary $p$, as well as \eref{peq.2} for  $p=2$ and arbitrary $n$. Also, the first two terms in the above expression add up to a positive number for all $p>2$. This proves our statement that when tensoring theories with $p>2$, the value of $\ell$ always increases. 

In principle one can generalise tensor products to {\em orbifolds} of tensor products. This might enable us in principle to recover theories with small numbers of characters and low values of $\ell$ from other such theories, but we defer this for future work.

\section{Two-character differential equations}

We now turn to the formulation and analysis of modular-invariant differential equations for two-character CFT's. The most general such second-order equation is of the form:
\be
\Big(D^2 + \phi_1(\tau)D+\phi_0(\tau)\Big)\chi=0
\ee
where $D$ is the covariant derivative defined in \eref{covdev}. We want to determine the possible coefficient functions $\phi_i(\tau)$ for the lowest allowed values $\ell=0,2,3,4,5$. For $\ell=0$ this was done in Ref.\cite{Mathur:1988na} while the $\ell=2$ case was investigated in Ref.\cite{Naculich:1988xv} to a few orders. We will investigate $\ell=2,3$ in detail and will see that they are as tractable as $\ell=0$, while for $\ell\ge 4$ there is at least one additional parameter in the equation and the standard Diophantine analysis is not so straightforward.

Since $D$ carries a modular weight of 2, $\phi_1$ must have modular weight 2 while $\phi_0$ has modular weight 4. If $\ell=0$ they are both non-singular, and it follows immediately that $\phi_1=0, \phi_0\sim E_4(\tau)$. 
The 2nd order equation with $\ell=0$ is therefore:
\be
(\tD^2 + \mu E_4)\chi=0
\label{leq.0}
\ee
where $\mu$ is a parameter. For convenience we have defined $\tD\equiv\frac{D}{2\pi i}$ which simplifies all the expressions. The above equation can be written more explicitly in terms of ordinary derivatives as:
\be
\left(\tdel^2 -\sfrac{1}{6}E_2\tdel + \mu E_4\right) \chi=0
\ee
where $\tdel\equiv \frac{\del}{2\pi i}$. 

Next consider $\ell=2$. In this case $\phi_0$ and $\phi_1$ can have a pole of maximum degree $\frac{1}{3}$ (the fractional degree means that the pole, if it occurs, must be located at $\tau=e^{\frac{i\pi}{3}}$ and counts as a $\frac16$-order pole). Now $E_4$ has a double zero at this point, while $E_6$ is nonvanishing. Thus we find $\phi_1\sim \frac{E_6}{E_4}$. On the other hand, there is no weight-4 expression that can be made from $E_4, E_6$ that has a single power of $E_4$ in the denominator. Hence we must have $\phi_0\sim E_4$. The $\ell=2$ equation is therefore:
\be
\left(\tD^2 + \mu_1\frac{E_6}{E_4}\tD+\mu_2 E_4\right)\chi=0
\label{leq.2}
\ee
It appears that we now have a two-parameter equation but, as we will see once we start analysing it, $\mu_1$ is completely determined given the order of the equation and the value of $\ell$. Thus we will only have to scan over one real parameter.

For $\ell=3$, the only possibility is a zero at the point $\tau=i$, which counts as a $\half$-order zero. Thus we cannot have any power of $E_4$ in the denominator. However $E_6$ vanishes at the point $\tau=i$, where it has a zero of degree $\half$, corresponding precisely to $\ell=3$. Thus $\phi_k$ is allowed to have $E_6$ in the denominator. It follows that $\phi_1\sim \frac{E_4^2}{E_6}$. On the other hand there is no modular object of weight 4 with $E_6$ in the denominator. Thus again, $\phi_0\sim E_4$. The $\ell=3$ equation is then:
\be
\left(\tD^2 + \mu_1\frac{E_8}{E_6}\tD+\mu_2 E_4\right)\chi=0
\label{leq.3}
\ee
As in the previous case, it turns out that $\mu_1$ is determined. 

For $\ell=4$, we can allow $\phi_k$ with at most $E_4^2$ in the denominator. Modular weight considerations then lead to $\phi_1\sim \frac{E_6}{E_4}$ and $\phi_0\sim \frac{E_6^2}{E_4^2}, E_4$. The $\ell=4$ equation is thus:
\be
\left(\tD^2 + \mu_1\frac{E_6}{E_4}\tD+\mu_2 E_4+\mu_3 \frac{E_6^2}{E_4^2}\right)\chi=0
\label{leq.4}
\ee
We see that the equation now has three parameters. After fixing one of them as before, we will find two free parameters to be scanned. 

The equation for $\ell=5$ is a little more subtle. We can allow $E_4, E_6$ and $E_4 E_6$ in the denominator, however there is no expression of weights 2 or 4 that achieves the last one. The most general equation is therefore:
\be
\left(\tD^2 + \left(\mu_1\frac{E_6}{E_4}+\mu_2\frac{E_4^2}{E_6}\right)\tD+\mu_3 E_4\right)\chi=0
\label{leq.5}
\ee
For $\ell=6$ we can now admit a full zero in the interior of moduli space (for $\ell\le 5$ we only had fractional zeroes confined to the orbifold points on the boundary). And for large $\ell$ the number of parameters grows as $\sim \ell^2$. Some interesting observations can be made about these cases, but in this part of the discussion we restrict ourselves to the lowest-lying cases $\ell\le 5$ and return to general $\ell$ at the end.

\section{Analysis of the solutions}

\subsection{$\ell=0$}

To find solutions of \eref{leq.0} that are potential CFT's, insert the series expansion $\chi=\sum_{n=0}^\infty a_n q^{\alpha+n}$ and $E_a(\tau)=\sum_{k=0}^\infty E_{a,k}\,q^k$ into the equation. This leads to:
\be
(n+\alpha)^2 a_n - \frac16 \sum_{k=0}^n (n-k+\alpha)E_{2,k}a_{n-k} + \mu \sum_{k=0}^n E_{4,k}a_{n-k}=0
\ee
The $n=0$ equation is the ``indicial equation'':
\be
\alpha^2-\sfrac16\alpha+\mu=0
\ee
Denote the roots of this equation by $\alpha_0,\alpha_1$ in increasing order, then:
\be
\begin{split}
\alpha_0+\alpha_1&=\frac16\\
\mu = \alpha_0\alpha_1 &=\alpha_0\left(\frac16-\alpha_0\right)
\end{split}
\label{muval}
\ee

Now we can use the $n=1$ equation to get:
\be
m_1\equiv\frac{a_1}{a_0}=-\frac{24\alpha+1440\mu}{5+12\alpha}
\label{neqone}
\ee
The two values of $\alpha_0,\alpha_1$ give rise to the corresponding quantities $m_1^{(0)}$ and $m_1^{(1)}$. In the former case this is the degeneracy of the first excited state in the identity character, while in the latter it is the ratio of the degeneracies of the first excited state and the ground state for the character of a nontrivial primary. Since $\mu$ is symmetric in $\alpha_0,\alpha_1$ we can combine Eqs.(\ref{muval}),(\ref{neqone}) and write:
\be
m_1^{(i)}=\frac{24\alpha_i(60\alpha_i-11)}{5+12\alpha_i}
\label{m.1.ell.0}
\ee
for $i=0,1$. As a check, the SU(2) WZW model at $k=1$ has two characters with $\alpha_0=-\frac{1}{24}$ and $\alpha_1=\frac{5}{24}$. The above equation then gives $m_1^{(0)}=3$ and $m_1^{(1)}=1$. The first value is the dimension of SU(2) while the second value, along with the fact that the primary is doubly degenerate (being spin-$\half$) tells us that there are two excited states above this primary. 

We can find some more constraints that turn out to be special to two-character theories. Setting $i=0$ in the above equation and using $\alpha_0=-\frac{c}{24}$ we get:
\be
m_1^{(0)}= \frac{c(5c+22)}{10-c}
\ee
This places an upper bound $c<10$. Also, by a little manipulation we find (we temporarily drop the $(0)$ superscript on $m_1$ to simplify the notation):
\be
(5c)^2 + 5c(m_1+22)=50 m_1
\label{fivec}
\ee
from which it follows that $5c$ is an integer. To summarise, for theories with two characters and $\ell=0$ we have shown that $c$ is a multiple of $\frac15$ and bounded above by 10. 

Now let us find all possible CFT characters that satisfy this equation. For this, notice from \eref{fivec} that $5c$ will be rational only if the discriminant:
\be
\sqrt{m_1^2+ 244 m_1+484}
\ee
is rational. Since $m_1$ has to be an integer, this can only be so if the above square root is in fact an integer. Thus we have:
\be
m_1^2+244 m_1+484=N^2
\ee 
for some integer $N$. To solve this, we shift $N$ by an integer amount that absorbs the first two terms on the left. Thus:
\be
N={\tilde N}+m_1 + 122
\ee
Inserting this into the above equation we find:
\be
m_1=-\frac{\tN}{2}-122 -\frac{7200}{\tN}
\ee
This tells us that $\tN$ is a negative even integer with $0<|\tN|<120$, and it must divide 7200. Also $\tN$ and $\frac{14400}{\tN}$ give us the same $m_1$. This restricts $\tN$ to the 22 values:
\be
\tN=-\{2,4,6,8,10,12,16,18,20,24,30,32, 36, 40, 48, 50, 60, 72, 80,90, 96,100 \}
\ee
Each of these can be used to find the corresponding value of $c$ from \eref{fivec}. Thereafter one can compute $a_n^{(i)}$ for both $i=0$ and 1 and over a large range of $n$. 

The results were reported in Ref.\cite{Mathur:1988na}, where 10 pairs of potential CFT characters were found. A couple of subtleties arose there that could be relevant for our subsequent investigations. First of all, a one-character theory appeared in the set corresponding to the $E_8$ affine Lie algebra CFT at level $k=1$, with $c=8$. Second, a nonunitary minimal model appeared there. To discover this, one had to admit the possibility that $\alpha_0 = -\frac{c}{24}$ is the higher of the two $\alpha's$, this happens only when the nontrivial primary has a negative conformal dimension. On doing this, a non-unitary CFT with $c=-\frac{22}{5}$ and $h=-\frac15$ was found in the list. This is the famous Lee-Yang edge singularity CFT. This teaches us that we must keep our eyes open for possible non-unitary CFT's and whenever necessary, reinterpret the formulae accordingly.

We now turn to an analysis of cases with $\ell>0$.

\subsection{$\ell=2$}

Here we work with the equation:
\be
\left(\tD^2 + \mu_1\frac{E_6}{E_4}\tD+\mu_2 E_4\right)\chi=0
\ee
Reduced to ordinary derivatives, this reads:
\be
\left(\tdel^2 -\sfrac16 E_2\tdel+\mu_1\frac{E_6}{E_4}\tdel+\mu_2 E_4\right)\chi=0
\ee
It is useful at this stage to introduce the Ramanujan identities satisfied by the Eisenstein series:
\be
\begin{split}
\tD E_2&=-\sfrac{1}{12}E_4\\
\tD E_4&=-\sfrac13 E_6\\
\tD E_6&=-\sfrac12 E_4^2
\end{split}
\label{ramanujanid}
\ee
where $\tD f\equiv \tdel f-\frac{r}{12}E_2f$ and $r$ is the weight of $f$. These will be helpful in rewriting the recursion relations in a form linear in the Eisenstein series. Multiplying the above differential equation by $E_4$ and using the second Ramanujan identity in \eref{ramanujanid} to eliminate the product $E_2E_4$, we have:
\be
\left(E_4\tdel^2 -\shalf\tdel E_4\tdel +\left(\mu_1-\sfrac16\right)E_6\tdel +
\mu_2 E_8\right)\chi=0
\ee
Inserting the mode expansion, we find:
\be
\sum_{k=0}^n \Big[(n+\alpha-k)^2 E_{4,k} -\sfrac12  k(n+\alpha-k) E_{4,k}
+{\tilde\mu}_1  (n+\alpha-k) E_{6,k} +\mu_2 E_{8,k}\Big]a_{n-k}=0
\label{recur.2}
\ee
where ${\tilde\mu}_1=\mu_1-\frac16$.

The indicial equation arises for $n=0$:
\be
\alpha^2 + {\tilde\mu}_1\alpha +\mu_2=0
\ee
Now we employ the identity \eref{valenceform} with $p=2,\ell=2$ to get:
\be
\alpha_0+\alpha_1 = \frac16 -\frac13 = -\frac16
\ee
At the same time the indicial equation tells us that $\alpha_0+\alpha_1 = -{\tilde \mu}_1$. It follows that $\tilde \mu_1=\frac16$. Thus, as promised, there is only one free parameter, $\mu_2$, in the equation. Moreover, we learn that:
\be
\mu_2 = - \alpha^2 - \sfrac16 \alpha
\ee

The next step is to consider the $n=1$ equation, which leads to:
\be
m_1^{(i)}=\frac{a_1^{(i)}}{a_0^{(i)}}=\frac{24\alpha_i(71+60\alpha_i)}{7+12\alpha_i}
\label{mfromalpha.2}
\ee
Restricting to $i=0$, dropping the superscript on $m_1$ to simplify the notation, and assuming that $\alpha_0=-\frac{c}{24}$ (we temporarily ignore the caveat at the end of the previous subsection) we find:
\be
m_1 = \frac{c(5c-142)}{(14-c)}
\ee
We see that for $m_1\ge 0$ we must have $14<c\le \frac{142}{5}$. Moreover, writing the above equation as:
\be
(5c)^2 + 5c(m_1-142)=70 m_1
\ee
shows that again, $5c$ is an integer. Applying the same analysis as in the previous subsection, we have:
\be
m_1^2-4m_1+20164=N^2
\ee
where $N$ is an integer. Now we shift:
\be
N=\tN+m_1-2
\ee
As before, this has the effect of removing all terms in $m_1$ from the LHS, but it introduces a term $m_1\tN$. One can then divide out by $\tN$ to solve for $m_1$:
\be
m_1=-\frac{\tN}{2}+2+\frac{10080}{\tN}
\label{mfromtn.2}
\ee
This tells us that $\tN$ is an even integer that divides 10080, however it must be positive if $\tN\le 144$ and negative otherwise, to ensure that $m_1\ge 0$. Moreover the formula is invariant under $\tN\to -\frac{20160}{\tN}$. Thus we can restrict the possible values of $\tN$ to 
\be
\begin{split}
\tN=\{&2, 4, 6, 8, 10, 12, 14, 16, 18, 20, 24, 28, 30, 32, 36, 40, 42, 48, 56, 60, 70,\\ & 72, 80, 84, 90, 96,112, 120, 126, 140, 144\}
\end{split}
\label{ell.2.list}
\ee
This is a finite set of 31 possible theories.

For each of these values of $\tN$, one uses \eref{mfromtn.2} to compute the value of $m_1$. From the above analysis, this is guaranteed to be a non-negative integer. Recalling that we have been using $m_1$ as shorthand for $m_1^{(0)}$, we can compute $\alpha_0$ from \eref{mfromalpha.2}. Next we use $\alpha_1=-\frac16-\alpha_0$ to calculate $\alpha_1$, and from this using \eref{mfromalpha.2} again, we get $m_1^{(1)}$. Thus we have the values of the central charge, the conformal dimension of the nontrivial primary, and the ratio of the number of states at the first excited level relative to the ground state for each character. This is a relatively long list, in one-to-one correspondence with the 31 values of $\tN$ listed above. The only way to rule out any of them at this stage would be if $m_1^{(1)}$ is negative. Interestingly, none of the candidates in the above list is ruled out in this way. 

Next, we continue by computing $m_2^{(0)}$ from \eref{recur.2}. If this comes out to be negative or fractional, this immediately rules out the candidate theory. The values  $\tN=\{ 2,4,6,10,14,18,28,32\}$ lead to fractional values of $m_2^{(0)}$, while $\tN=\{126,140,144\}$ lead to negative values of $m_2^{(0)}$. $\tN=120$ gives rise to a divergence in the value of $m_2^{(0)}$. Thus all these candidates are ruled out, already at level $n=2$. At the next level $n=3$ we find no negative values, but we do get fractional values of $m_3^{(0)}$ for $\tN=\{8,36,56,70\}$. Thus these cases are also eliminated. At level 4, we get fractional $m_1^{(0)}$ for $\tN=\{12, 16, 42, 84,112\}$. At this stage we are left with just 10 theories, corresponding to:
\be
\tN=\{20, 24, 30, 40, 48, 60, 72, 80, 90, 96\}
\label{nearfinal}
\ee

The next step is to consider $m_n^{(1)}$. If this comes out negative then the candidate is not a consistent CFT. To start with, we examine these numbers up to $n=20$ and find that they are all non-negative. They are generically fractional and display two markedly different kinds of behaviour. For some candidate theories, $m_n^{(1)}$ is a fraction whose denominator is relatively small and contains only two or three prime factors. By the time we reach $n=10$ this denominator is ``stable'' in the sense that new denominator factors do not appear. In such cases we may assume the ground state has a degeneracy equal to the largest denominator encountered. This then leads to non-negative integer values for the degeneracies of all the excited states. However, in other cases the denominator keeps increasing steadily as we increase $n$. In this situation we must exclude the candidate. 

Among the list of 10 candidates in \eref{nearfinal}, precisely one has growing denominators by the time we reach the 10th level, namely $\tN=20$. We checked this case upto the 1000th level and found that the denominator has grown to be of order $10^{1250}$. It must be admitted that we do not know of a mathematically rigorous way to assure that the denominator does not stabilise at some higher level, but the evidence seems rather compelling that it does not. Moreover, we can easily explain this candidate theory, which has $c=16$ and $m_1=496$. At these values we know a single character that is modular invariant upto a phase, namely $\chi=j^{\frac23}$. This corresponds to the famous level$-1$ $E_{8}\times E_{8}$ and ${\rm Spin}(32)/Z_2$ CFT's. This character has $\ell=4$, but one can easily show that it also arises as one solution of a second-order $\ell=2$ differential equation whose second solution is a spurious character (an analogous situation arose in Ref.\cite{Mathur:1988na} for the $c=8$ case). Thus we can confidently exclude it from consideration here.

It is noteworthy that if we look at the behaviour of the nontrivial primary character for cases already ruled out by examining the identity character (i.e. cases not in the list of \eref{nearfinal}), then we find that many of them have increasing denominators. Thus the consistency criteria applied independently to the identity and the nontrivial primary characters seem to largely point to the same conclusions, although in the former case the criterion (non-negative integers) is rigorous, while in the latter case the criterion (non-growing denominators) is hard to make rigorous.

\begin{table}[ht]
\centering
\begin{tabular}{|c||c|c|c|c|c|}
\hline
No. & ${\tilde N}$ & $c$ & $h$ & $m_1$ & \small App. degen.\\
\hline
1 & 24 & $\frac{82}{5}$ & $\frac65$ & 410& 902 \\
\hline
2 & 30 & 17 & $\frac54$ & 323 & 51 \\ 
\hline
3 & 40 & 18 & $\frac43$ & 234 & 1  \\
\hline
4 & 48 & $\frac{94}{5}$ & $\frac75$ & 188 & 4794\\
\hline
5 & 60 & 20 & $\frac32$ &140 & 5 \\
\hline
6 & 72 & $\frac{106}{5}$ & $\frac85$ & 106 & 15847 \\
\hline
7 & 80 & 22 & $\frac53$ & 88 & 22 \\
\hline
8 & 90 & 23 & $\frac74$ & 69 & 253 \\
\hline
9 & 96 & $\frac{118}{5}$ & $\frac95$ & 59 & 32509 \\
\hline
\end{tabular}
\caption{Potentially consistent CFT's with $\ell=2$.}
\label{table-leq.2}
\end{table}

At the end we find a list of 9 potentially consistent sets of characters, shown in Table \ref{table-leq.2} together with the corresponding leading degeneracies, central charges and dimension of the nontrivial primary. The heading ``App.\,degen.'' in the last column denotes the apparent degeneracy of the nontrivial primary. This is only a lower bound -- the minimum number by which the character should be normalised such that the coefficients become integer. 

Table \ref{table-leq.2} was generated just by considering degeneracies upto $n\sim 10$. The acid test is now to check whether the degeneracy of the identity character $m_n^{(0}$ and of the nontrivial primary $m_n^{(1)}$ are non-negative integers (after choosing a fixed ground-state degeneracy in the latter case) upto a very large value of $n$. Remarkably, we find that every one of the above candidates satisfies both criteria upto $n=5000$. This stability is remarkable, and strongly suggests that these 9 theories correspond to conformal field theories. It can be seen from the table that their central charges lie in the range $16 < c < 24$. Some more details are provided in Appendix I, where details are provided for a successful candidate showing the positive integer values $m_1^{(0)}$ and positive rational $m_1^{(1)}$ with stable denominators. Another example displayed there is of an unsuccessful candidate where the $m_1^{(0)}$ include negative integers and the $m_1^{(1)}$ are rational numbers with rapidly increasing denominators. 

It is noteworthy that the table above closely matches that for $\ell=0$ theories with two characters, first found in Ref.\cite{Mathur:1988na}. There is a close similarity between the spectrum of central charges and conformal dimensions. This is examined at greater length in Section 6 below.

In Ref.\cite{Naculich:1988xv}, a class of characters with $\ell=2$ was found by a different and powerful method (independently discussed in Ref.\cite{Mathur:1988gt}). First, the original $\ell=0$ differential equation is mapped to a Fuchsian equation in the Picard $\lambda$-function, in terms of which explicit solutions can be found for the characters in terms of hypergeometric functions. These are then examined directly to check positive integrality of the coefficients in $q$. Taking this idea further, Ref.\cite{Naculich:1988xv} proposed a general form for the characters by treating the $j$-function, or rather its inverse $g(\tau)=\frac{1728}{j(\tau)}$, as the independent variable. By looking at the coefficients at low values, a list of potential characters with $\ell=2,8,10$ were found. The central charges and conformal dimensions in that paper for $\ell=2$ are in perfect agreement with our computations. The additional power of the method of Ref.\cite{Naculich:1988xv}, however, is that it directly provides the degeneracy of the non-identity character while as we indicated, our numbers are only lower bounds. The actual degeneracies computed in Ref.\cite{Naculich:1988xv} do not all match the entries in our table, being much larger in many cases, but are divisible by them.

Another work related to our discussion is Ref.\cite{Kiritsis:1988kq}. Here the same Fuchsian differential equation method is used, but only for $\ell=0$, and the condition of positivity of coefficients is relaxed. Looking at all the central charges in Table \ref{table-leq.2} of that paper, we find that after shifting the central charges by 4, the 12th to 20th entries of Table I of that paper can be matched with the entries of our table above. Moreover, the conformal dimensions of the nontrivial primary also match. However, the degeneracies of the nontrivial primary do not match at all -- for example our theory $c=\frac{106}{5}$ and $h=\frac85$ can be matched with the entry with $c=\frac{86}{5}$ and $h=\frac85$ in Ref.\cite{Kiritsis:1988kq}, but we find the primary to be 15847-fold degenerate while that reference finds the degeneracy to be 12857. More significant is that one cannot perform a shift of $c$ by 4 using any of the standard functions of $j$, which is why the analysis of Ref.\cite{Kiritsis:1988kq} admits shifts only by multiples of 8 or 12. It appears therefore that the results of Ref.\cite{Kiritsis:1988kq} cannot be literally correct, in fact if true then they would exclude the characters in our Table \ref{table-leq.2}. Nevertheless, it is a potentially useful idea to try and classify all $\ell>0$ theories in terms of those with $\ell=0$ and we hope to address this question in the future.

\subsection{$\ell=3$}

Here we work with the equation:
\be
\left(\tD^2 + \mu_1\frac{E_8}{E_6}\tD+\mu_2 E_4\right)\chi=0
\ee
In terms of ordinary derivatives, this can be written:
\be
\left(\tdel^2 -\frac16 E_2\tdel+\mu_1\frac{E_8}{E_6}\tdel+\mu_2 E_4\right)\chi=0
\ee
Multiplying through by $E_6$, using the Ramanujan identity \eref{ramanujanid} to eliminate $E_2E_6$,  and inserting the mode expansion, we have:
\be
\sum_{k=0}^n \Big[(n+\alpha-k)^2 E_{6,k} -\sfrac13  k(n+\alpha-k) E_{6,k}
+{\tilde\mu}_1  (n+\alpha-k) E_{8,k} +\mu_2 E_{10,k}\Big]a_{n-k}=0
\ee
where ${\tilde\mu}_1=\mu_1-\frac16$.

The indicial equation arises for $n=0$:
\be
\alpha^2 + {\tilde\mu}_1\alpha +\mu_2=0
\ee
Now we employ the identity \eref{valenceform} with $p=2,\ell=3$ to get:
\be
\alpha_0+\alpha_1 = \frac16 -\half = -\frac13
\ee
At the same time the indicial equation tells us that $\alpha_0+\alpha_1 = -{\tilde \mu}_1$. It follows that $\tilde \mu_1=\frac13$. Thus, as promised, there is only one free parameter ($\mu_2$) in the equation.

The next step is to solve for $m_1$ in terms of $\alpha$. We find:
\be
m_1^{(i)}=\frac{24\alpha^{(i)}(15\alpha^{(i)}-26)}{2+3\alpha^{(i)}}
\ee
With $i=0$ and $\alpha^{(0)}=-\frac{c}{24}$, we get:
\be
(5c)^2+(m_1+208)(5c)=80m_1
\ee
This again shows that $5c$ is an integer.

Following the steps outlined previously, we end up with:
\be
m_1=-\frac{\tN}{2}-\frac{46080}{\tN}-368
\ee
This tells us that $\tN$ has to be a negative even integer that divides 46080. We can restrict to the range $|\tN|\le 160$ because after this $m_1$ becomes negative until $\tN=-576$. After that it turns positive again, but this range is related to the first one by the symmetry $\tN\to \frac{92160}{\tN}$. It follows that the allowed values of $\tN$ are: 
\be
\tN=-\{ 
2, 4, 6, 8, 10, 12, 16, 18, 20, 24, 30, 32, 36, 40, 48, 60, 64, 72, 80, 90, 96, 120, 128, 144, 160\}
\ee

We now subject these cases to the same tests as in the previous subsection ($\ell=2$) and find a strikingly different result. At the first level, we find that $m_1^{(1)}$ is negative for a number of the above cases. This already rules out all candidates except $\tN = -\{80,90,96,120,128,144,160\}$. So we are already down to 7 candidates. Next we examine the first few $m_n^{(0)}$. At level 2, $\tN=-128$ and $\tN=-144$ are eliminated (fractional values). At level 3, $\tN=-90$ and $\tN=-96$ are ruled out in the same way. Thus we are left with three candidates, with central charge assignments $c=0,4,8$. Of these, the $c=4$ case gives $h=0$ and hence the two characters are the same. This seems unphysical, and moreover there is no single-character CFT with $c=4$. The case with $c=8$ has $\alpha_1=0$. From the recursion relations we have seen that every $m_n^{(i)}$ is proportional to $\alpha_i$. Thus all the infinitely many $m_n^{(1)}$ vanish and the second character has no tower associated to it. It is easy to see that the first character is nothing but $j(\tau)^{\frac13}$. Thus we have rediscovered the $E_8,k=1$ single-character theory. The last case is the theory apparently having $c=0$. In this case we find $\alpha_0=0$, so the putative identity character has no tower. On the other hand, $\alpha_1=-\frac13$, so what we thought of as the non-identity character is really $j(\tau)^\frac13$. Thus, once more we find the $E_8,k=1$ theory. 

We conclude that there are no new two-character CFT's with $\ell=3$. 
It is amusing to note that there are also no 1-character CFT's with $\ell=3$. The only candidate for this would have been $(j-1728)^\half$, but this unfortunately contains negative coefficients in the power-series expansion. However we do know of at least one CFT with $\ell=3$. As mentioned at the end of Section 2, this is the $Z_2$ orbifold of a free boson at radius 2 (corresponding to $p=2$) for which there are 6 characters.

\subsection{$\ell=4$}

Here, for the first time, we encounter an equation where there are more free parameters than roots. In terms of ordinary  derivatives, and after multiplying throughout by $E_4^2$, \eref{leq.4} can be expressed:
\be
\left(E_8 \tdel^2-\sfrac16 E_2 E_4^2 \tdel +\mu_1 E_{10} \tdel+\nu E_4^3  + \mu\Delta\right)\chi=0 
\ee
Here we have made use of the fact that modular forms of weight 12 are proportional to an arbitrary linear combination of $E_4^3$ and $E_6^2$, and have chosen one of the linear combinations to be $\Delta$ defined in \eref{jdeltadef}. This is convenient because $\Delta$ has no  constant term in the $q$-expansion. The constants have been relabelled to $(\mu_1,\nu,\mu)$ for convenience.
 
To write it as a linear equation in the Eisenstein series, we again use the Ramanujan identities \eref{ramanujanid} to get:
\be
\left(E_8 \tdel^2-\sfrac14 \tdel E_8 \tdel +(\mu_1-\sfrac16) E_{10}\tdel+\nu E_4^3 +\mu \Delta\right)\chi=0 
\ee
In terms of modes, this is:
\be
\sum_{k=0}^n \Bigg[(n+\alpha-k)^2 E_{8,k}+\Big((\mu_1-\sfrac16)E_{10,k}-\sfrac14 kE_{8,k}\Big)(n+\alpha-k)+\Big(\nu (E_4^3)_{,k} + \mu \Delta_{,k}\Big)\Bigg]a_{n-k}=0
\label{recur.4}
\ee
The indicial equation is:
\be
\alpha^2 + (\mu_1-\sfrac16)\alpha+ \nu=0
\ee
Comparing with \eref{valenceform} with $\ell=4$, we see that $\mu_1=\frac23$. Then we have:
\be
\begin{split}
\alpha_0+\alpha_1 &= -\half\\
\alpha_0\alpha_1 &= \nu
\end{split}
\ee
Next, considering the recursion relation \eref{recur.4} for $n=1$, we find:
\be
m_1 = \frac{24\alpha(20\alpha+51)}{4\alpha+3}-\frac{2\mu}{4\alpha+3}
\label{m.1.ell.4}
\ee
The presence of the extra parameter $\mu$ means that we must now look for a pair $(\alpha,\mu)$ that makes $m_1$ a non-negative integer. As we will see, this is an infinite list. 
So let us start by making a number of general statements. First of all, if $\alpha$ is rational and $m_1$ integer then the above equation tells us $\mu$ is rational. Second, if we find any pair $(\alpha,\mu)$ that makes $m_1$ an integer, then we can always shift $\mu\to \mu-n\frac{4\alpha+3}{2}$ where $n$ is an arbitrary integer, which has the effect of shifting $m_1\to m_1+n$. This in particular generates an infinite set of possibilities. This means the approach that was followed for $\ell=0,2,3$ will not work here. 

We will see below that a Diophantine equation emerges if we also require $m_2$ to be a non-negative integer. But first let us consider the discriminant obtained as in the previous sections, which this time comes out to be:
\be
93636 + 240 \mu -252 m_1 + m_1^2
\ee
Since $\mu$ need not be integer, we can no longer impose that the above is a square of an integer $N$. To consider one example, let us arbitrarily choose $\mu=0$. In this case the previous type of analysis applies, and we find:
\be
m_1 = \frac{38880}{\tN} + 126 - \frac{\tN}{2}
\ee
Therefore $\tN$ must be an even integer that divides 38880. We also have an equivalence under $\tN\to -\frac{77760}{\tN}$. Thus the list of possible values is:
\be
\begin{split}
\tN=\{ &2, 4, 6, 8, 10, 12, 16, 18, 20, 24, 30, 32, 36, 40, 48, 54, 60, 72, 80, 90,\\ 
& 96, 108, 120, 144, 160, 162, 180, 216, 240, 270, 288, 324, 360, 432\}
\end{split}
\ee
Going to level 2, we rule out all the above integers except for:
\be
\tN=\{12, 20, 24, 30, 40, 48, 60, 72, 80, 90, 96, 108\}
\label{ruleout}
\ee
Of these, $\tN=12,20,108$ get ruled out at level 3 and $\tN=30,48,96$ are eliminated at level 4. The remaining values: $\tN=24,40,60,72,80,90$ get ruled out by the time we reach level 10. We conclude there are no $\ell=4$ theories with $\mu=0$.

The choice $\mu=0$ was, of course, arbitrary. Indeed the negative result above supports the notion that characters satisfying the conditions to describe CFT only occur in very special places. We can exhibit at least one interesting choice for $\mu$ for which CFT's are known to exist. To see this, recall that if we tensor any of the 2-character $\ell=0$ theories having a given $\alpha_0$ with the 1-character theory whose character is $j^\frac13$, namely the $E_8$ theory at $k=1$, we end up with a 2-character theory with $\ell=4$. This tensored theory has a central charge shifted by 8 from the original 2-character theory, which means $\alpha_0$ is replaced by $\alpha_0-\frac13$. Also the original first-level degeneracy $m_1$ is augmented by 248, the first-level states contributed by the $E_8$ character. Thus if we take \eref{m.1.ell.4} and make the replacement $m_1\to m_1+248$ and $\alpha_0\to \alpha_0-\frac13$, there must be some value of $\mu$ for which this equation reduces to \eref{m.1.ell.0}. This turns out to be the case for $\mu=-384$, independent of $\alpha$.

Returning now to general $\mu$, we can use the recursion relation with $n=2$ to find:
\be
m_2 = \frac{2(264 \mu + \mu^2 + 142938 x - 216 \mu x + 284976 x^2 - 480 \mu x^2 + 
 298080 x^3 + 57600 x^4)}{(3 + 4 x) (5 + 4 x)}
 \ee
Eliminating $\mu$ between this and \eref{m.1.ell.4} we get:
\be
m_2= \frac{406008 x + 246240 x^2 - 528 m_1 - 2016 x m_1 + 3 m_1^2 + 
  4 x m_1^2}{2 (5 + 4 x)}
  \ee
Unlike the preceding equation, this is quadratic in $x$ and the coefficients are determined in terms of non-negative integers $m_1,m_2$. Thus the discriminant must be the square of a non-negative integer:
\be
(-406008 + 2016 m_1 - 4 m_1^2 + 8 m_2)^2 + 
 984960 (528 m_1 - 3 m_1^2 + 10 m_2)=N^2
  \ee
In this way we have achieved a Diophantine equation. We cannot show as before that this has finitely many solutions, indeed in this case there may well be infinitely many solutions. But at least they are now labelled by integers. This is an (admittedly slight) improvement over the previous equation which was labelled by an arbitrary rational number $\mu$. 

We will not pursue the case of $\ell=5$ here, because the considerations are similar to $\ell=4$. The situation for $\ell=4,5$ is in fact quite similar to that of three-character theories with $\ell=0$. In both cases there is a differential equation labelled by two parameters (in the former case, $\alpha_0$ and $\mu$, while in the latter, $\alpha_0$ and $\alpha_1$). However in the latter case we know more: as pointed out in Ref.\cite{Mathur:1988gt}, there are several infinite families of three-character CFT's with $\ell=0$, including the SO(N) WZW models at level $k=1$ for arbitrary $N$. We hope to return in future to a discussion of this family of differential equations. 

\section{Chiral algebras for the $\ell=2$ characters}

\subsection{Identifying potential current algebras}

As we saw above, there appears to be a set of 9 two-character CFT's with $\ell=2$. These originally appeared in Ref.\cite{Naculich:1988xv} and we have confirmed that they satisfy the desired properties to high orders in the $q$-expansion. However to date, there has been no analysis of these theories in terms of identifying their chiral algebras and help us understand what precisely they are, if indeed they are completely consistent as CFT's. In this section we take a step in this direction, with encouraging results.

Before going ahead with this analysis, let us note a striking resemblance between certain aspects of the $\ell=2$ characters in Table \ref{table-leq.2} of the present paper and the $\ell=0$ characters in Table 1 of Ref.\cite{Mathur:1988na}. Ignoring the last entry in the latter paper which is the one-character $E_8$ theory, we have precisely 9 entries in each table. There is an evident relation between the central charges, namely that those in the former table are obtained by adding 16 to those in the latter table. Moreover the conformal dimensions of the nontrivial primary in the former table are obtained by adding 1 to those in the latter table. On the other hand, there is no obvious relation between the $m_1$ in the two tables, nor between the degeneracies of the non-trivial primaries. There is no way to explain the observed relation by tensoring the $\ell=0$ characters with those of a 1-character theory (functions of $j$). To see this, note that in order to have a shift of $c$ by 16, one would have to postulate that $\tchi_i = j^\frac23 \chi_i$ where $\chi_i$ are the $\ell=0$ characters and $\tchi_i$ are the $\ell=2$ characters. However this immediately means that the theory whose characters are $\tchi_i$ has $\ell=4$ rather than $\ell=2$. Moreover such a simple relation would mean that ${\tilde m}_1 =m_1+ 496$ and that the ground-state degeneracy of $\tchi_1$ is the same as that of $\chi_1$. However, neither of these is the case. Thus the $\ell=2$ theories, if they are completely consistent with the axioms of conformal field theory, have to be independent CFT's and not tensor products. 

Let us now proceed to analyse the current algebras of these theories. To this end we use the proposal in Section 6 of Ref.\cite{Mathur:1988gt} on how to identify the current algebra for a given set of characters. The idea is to consider a CFT whose chiral algebra is the sum of a number of simple Kac-Moody algebras: $G=\oplus_i G_i$. We start with the Sugawara relation:
\be
c_i=\frac{k_i \dG_i}{k_i+(c_v)_i}
\ee
where $k_i$ is the level of the $i$th algebra, $\dG_i$ is the dimension of the associated finite-dimensional Lie algebra and $(c_v)_i$ is the dual Coxeter number of this algebra. Now let us try to guess a combination of Kac-Moody algebras that underlie a given CFT when we know the central charge $c$ of that CFT as well as the integer $m_1$, the number of states at first level above the identity character. Since all such states are of the form $J_{-1}^a |0\rangle$ where $|0\rangle$ is the vacuum state, $m_1$ counts the total number of Kac-Moody currents. It follows\footnote{There is an assumption that the entire central charge of the theory comes from the current algebra. We thank Matthias Gaberdiel for pointing this out.} that:
\be
\begin{split}
c&=\sum_i \frac{k_i \dG_i}{k_i+(c_v)_i}\\
m_1 &= \sum_i \dG_i
\end{split}
\ee
While the above equations allow for different constituent algebras with arbitrary $k_i$ and $(c_v)_i$, there is a stronger result in Eq.(6.27) of Ref.\cite{Mathur:1988na}. This states that for two-character theories\footnote{The derivation there makes use of $\ell=0$ but can easily be generalised for any $\ell$.}, each constituent algebra separately satisfies:
\be
\frac{(c_v)_i}{k_i}=\frac{m_1}{c}-1
\ee
Thus, we only need to look for factors that all have the same value of $\frac{c_v}{k}$. Let us apply this case-by-case to the theories listed in Table \ref{table-leq.2}. For simplicity we restrict the search to $k_i=1$ although there could well be more solutions at higher levels (and we will provide one such example later). The goal at this stage is to see whether there exist any possible current algebras that can potentially arise in our $\ell=2$ theories. Even at level-1 we will find one or more solutions in every case, except the first and last.

We first list out all the algebras whose value of $c_v$ equals $\frac{m_1}{c}-1$. Next we identify those whose dimensions, with positive integral coefficients, add up to $m_1$. The considerations above ensure that the central charges likewise add up to the correct value. The Lie algebras with the right $c_v$, as well as the combinations of these that can reproduce the required $c$ and $m_1$ (after eliminating the standard equivalences at low ranks) are shown in Table \ref{table-cv}.

\begin{table}[ht]
\centering
\begin{tabular}{|c||c|c|c|c|c|}
\hline
No. & $c$ & $c_v=$ & Algebras & Combinations \\[-2mm]
& & $\frac{m_1}{c}-1$ & with given $c_v$ & with right $c,m_1$\\
\hline
1 & $\frac{82}{5}$ & 24 & $A_{23}, C_{23}, D_{13}$ & None\\
\hline
2 & 17 & 18 & $A_{17}, C_{17}, D_{10}, E_7$ & $A_{17},D_{10}\oplus E_7$\\ 
\hline
3 & 18 & 12 & $A_{11}, C_{11}, D_7, E_6$ & $A_{11}\oplus D_7, (E_6)^3$ \\
\hline
4 & $\frac{94}{5}$ & 9 & $A_8,B_5,C_8,F_4$ & $C_8\oplus F_4$\\
\hline
5 & 20 &6 & $A_5,C_5,D_4$ & $(A_5)^4, (D_4)^5$\\
\hline
6 & $\frac{106}{5}$ & 4 & $A_3,C_3,G_2$ & $A_3\oplus C_3\oplus (G_2)^5, A_3\oplus (C_3)^3\oplus (G_2)^2$\\
\hline
7 & 22 & 3 & $A_2,B_2$ & $(A_2)^{11}, (A_2)^6\oplus (B_2)^4,A_2\oplus (B_2)^8$\\
\hline
8 & 23 & 2 & $A_1$& $(A_1)^{23}$\\
\hline
9 & $\frac{118}{5}$ & $\frac52$ & None & None \\
\hline
\end{tabular}
\caption{Potential level-1 chiral algebras for the $\ell=2$ theories.}
\label{table-cv}
\end{table}

Table \ref{table-cv} shows that there are one or more candidates in all cases, except the first and last one. Now it happens that one can rule out the first and last theory anyway, as follows. It was shown in Refs.\cite{Mathur:1988na},\cite{Mathur:1988gt}  that the $\ell=0$ theories with $c=\frac{2}{5}, \frac{38}{5}$ are both inconsistent as they stand, because the fusion rule coefficients \cite{Verlinde:1988sn} calculated for these characters are not all positive integers. Instead, one of them turns out to be equal to $-1$. The first of these theories could be rescued by interchanging the role of the identity and non-identity characters, whereupon it became a non-unitary theory. This was possible because in that case {\em both} characters had non-degenerate ground states. The same interchange is not possible for the theory with $c=\frac{38}{5}$ because the nontrivial primary has a 57-fold degeneracy and therefore one cannot interchange it with the identity character \cite{Mathur:1988gt}. If now we consider the corresponding $\ell=2$ theories with central charges $\frac{82}{5},\frac{118}{5}$, we find that in both cases the nontrivial primary has a large degeneracy (at least 902 and 32509 respectively, as we see from Table \ref{table-leq.2}). Moreover, it has been shown in Ref.\cite{Naculich:1988xv} that this pair has the same modular transformation matrices (and hence fusion rules, from the Verlinde formula Ref.\cite{Verlinde:1988sn}) as the corresponding $\ell=0$ pair. It follows that they suffer from a negative fusion-rule coefficient and are not consistent CFT's. The fact that we cannot find even one potential Kac-Moody algebra for them is in agreement with this.

This leaves 7 theories, corresponding to entries $2-8$ of Table \ref{table-cv}, for which we can take the analysis further. It is rather remarkable that there are any candidate Lie algebras at all which satisfy the rather stringent numerological requirements that have been applied. Theories 4 and 8 with central charges $\frac{94}{5}$ and 23 each have a unique candidate algebra at level $k=1$. The remaining cases have multiple candidates. If we allow levels $k>1$ we could find more solutions, for example Theory 8 can also be matched with $(A_1)^3$ at level 1 combined with $(A_3)^4$ at level 2.

Note that although we have identified one or more potential Kac-Moody algebras that are symmetries of the $\ell=2$ theories, these theories are certainly not the WZW models of those algebras. For example if we simply take 23 copies of $A_1$ at level 1, the resulting theory with $c=23$ has 24 characters and from \eref{peq.2} it has $\ell=0$. This is quite different from the desired theory in our list, which of course also has $c=23$ but just two characters and $\ell=2$. 
We will leave for the future the question of completely classifying the possible Kac-Moody algebras for these characters, as well as of deriving the corresponding theories from those algebras.

\subsection{Some basic consistency conditions}

In order to support our conjectured identification of the current algebras for the theories in Table \ref{table-cv}, 
we focus attention on the nontrivial primary. Given that our $\ell=2$ theories are not simply the WZW models of those algebras, it must yet be true that the single primary of our theories arises as a subset of the primaries of those (in general tensored) algebras, and our characters must be a combination of those characters. Running through theories $2-8$ of Table \ref{table-leq.2}, we now ask in each case whether the conformal dimension of the nontrivial primary can arise as a combination of level-1 primaries of the algebras listed in Table \ref{table-cv}. The theories based on these current algebras will always have additional primaries which have to be eliminated. This would correspond to taking some suitable non-diagonal combination of the characters. We postpone for the future the goal of finding this combination in each case, and will content ourselves with the basic test of identifying the desired primary within the representations of each possible current algebra. The results are as follows.

\vspace*{2mm}

\noindent \underline{Theory 2, $c=17$}

\vspace*{2mm}

The primary in this theory has dimension $\frac54$. The candidate algebras here are $A_{17}$ and $D_{10}\oplus E_7$. The primaries of $A_{17}$ at level 1 have dimensions $h_r=\frac{r(18-r)}{36}$ with $r=1,2,\cdots 9$. The choice $r=3$ gives $h=\frac54$ as desired. This is the representation with three boxes in a single vertical column. On the other hand, $D_{10}$ at level 1 is a 3-character theory whose two nontrivial primaries have dimensions $\frac54$ and $\half$, and $E_7$ at level one has a single primary of dimension $\frac34$. Thus in the sum $D_{10}\oplus E_7$, there is indeed a primary of dimension $\frac54$ obtained by taking this primary of $D_{10}$ and tensoring with the identity of $E_7$. As expected, in each case there are other primaries that one has to somehow eliminate.

\vspace*{2mm}

\noindent \underline{Theory 3, $c=18$}

\vspace*{2mm}

The primary of this theory has dimension $\frac43$. The possible algebras are $A_{11}\oplus D_7$ and $(E_6)^3$. The primaries of $A_{11}$ have dimension $\frac{r(12-r)}{24}$, so $r=4$ gives us a primary of dimension $\frac43$ as desired, when tensored with the identity of $D_7$. There are also other possibilities: the $r=2$ primary of $A_{11}$ tensored with the $h=\half$ primary of $D_7$ also has dimension $\frac43$, and so does the $r=1$ primary of $A_{11}$ tensored with the $h=\frac78$ primary of $D_7$.  If instead we consider $(E_6)^3$, then this has a primary of dimension $\frac23$. Tensoring two $E_6$ primaries with one $E_6$ identity, which can be done in 3 ways, also generates a field of dimension $\frac43$. Thus both $A_{11}\oplus D_7$ and $(E_6)^3$ are viable level-1 current algebras for this set of characters.

\vspace*{2mm}

\noindent \underline{Theory 4, $c=\frac{94}{5}$}

\vspace*{2mm}

This theory has a primary with $h=\frac75$. There is a unique candidate algebra, $C_8\oplus F_4$. The $F_4$ algebra has a primary of dimension $\frac35$, while $C_8=Sp(16)$ at level 1 has primaries of dimension:
\be
\frac{17}{40}, \frac{4}{5}, \frac98, \frac75, \frac{13}{10}, \frac95, \frac{77}{40}, 2
\ee 
Thus the primary of Theory 4 can arise as the tensor product of the $h=\frac35$ primary of $F_4$ and the $h=\frac45$ primary of $C_8$, as well as the product of the identity of $F_4$ and the $h=\frac75$ primary of $C_8$. 

\vspace*{2mm}

\noindent \underline{Theory 5, $c=20$}

\vspace*{2mm}

The primary of this theory has $h=\frac32$. The possible current algebras are $(A_5)^4$ and $(D_4)^5$. $A_5$ at level 1 has primaries of dimension $\frac{r(6-r)}{12}=\frac{5}{12},\frac23, \frac34$. Thus when taking the fourth power, we can tensor the identity with the $h=\frac34$ primary twice each, to realise $h=\frac32$. For $D_4$, we only have primaries of level $\half$ (three distinct ones with the same character, because of triality). When taking the 5th power, we should keep the identity twice and the $h=\half$ primary three times to realise $h=\frac32$. Thus, again both candidate algebras are capable of producing the right primary.

\vspace*{2mm}

\noindent \underline{Theory 6, $c=\frac{106}{5}$}

\vspace*{2mm}

This theory has a primary with $h=\frac85$. The possible current algebras are $A_3\oplus$\,$C_3\oplus$ $(G_2)^5$ and 
$A_3\oplus (C_3)^3\oplus (G_2)^2$. Now $A_3$ has primaries of dimension $\frac38,\half$, while $C_3$ has primaries of dimension $\frac{7}{20},\frac35,\frac34$ and $G_2$ has a primary of dimension $\frac25$. Thus in the first case the only way to recover the desired primary is to take the $G_2$ primary four times, tensored with the identities of the other algebras. In the second case the power of $G_2$ is less than 4, however we are saved by the option of taking the $h=\frac35$ primary of $C_3$ twice, along with one copy of the $h=\frac25$ primary of $G_2$. It is noteworthy that although the same simple-algebra factors occur in both cases, there are different ways of getting the desired primary.

\vspace*{2mm}

\noindent \underline{Theory 7, $c=22$}

\vspace*{2mm}

In this case the desired primary has $h=\frac53$. The possible chiral algebras are $(A_2)^{11}$, $(A_2)^6\oplus (B_2)^4$ and $A_2\oplus (B_2)^8$. Now $A_2$ has a single (complex) primary of dimension $\frac13$, while $B_2$ has primaries of dimension $\half$ and $\frac{5}{16}$. We see that in the first two cases, the desired primary is found as five copies of the $A_2$ primary. However this is no longer an option in the last case and there is no other way to generate $h=\frac53$ in that case. Therefore we seem to have ruled out the possibility of $A_2\oplus (B_2)^8$ arising as a chiral algebra of this theory, while the first two possibilities are consistent at this stage.

\vspace*{2mm}

\noindent \underline{Theory 8, $c=23$}

\vspace*{2mm}

In this case the unique possible chiral algebra is $(A_1)^{23}$. We need to reproduce a primary with $h=\frac74$ and the building blocks are $A_1$ primaries of $h=\frac14$. Thus we just have to tensor the latter with itself 7 times, and with the $A_1$ identity 16 times.

To summarise this section, the successes encountered so far provide strong encouragement that the $\ell=2$ characters do indeed describe consistent, novel two-character CFT's.

\section{Conclusions}

We have revisited the classification of all RCFT's with two characters that satisfy a one-parameter differential equation. This case arises only when the number of zeroes $\ell$ of the Wronskian of the characters is $0,2,3$. For $\ell=2$ we verify the existence of 9 candidate CFT's, all with central charges in the range $16< c< 24$, in agreement with a prediction of Ref.\cite{Naculich:1988xv}. Further analysis is required to verify if these are fully consistent CFT's, for example their correlators satisfy the consistency conditions of Ref.\cite{Moore:1988uz}. For $\ell=3$ we were able to completely rule out the existence of any CFT's. Perhaps the most tantalising new result is the identification in Section 6 of current algebras that appear to govern the $\ell=2$ CFT's, which deserves to be explored further.

Now that it has been described in detail, let us briefly summarise our approach. First, obtain the recursion relation from the differential equation and write down the cases $n=0,1$. The first one gives rise to the indicial equation. The sum of roots is determined by the Riemann-Roch identity, \eref{valenceform}, hence only one of the exponents (which we take to be $\alpha_0=-\frac{c}{24}$) is independent. Now the indicial equation relates the product of roots to the free parameter in the equation, hence the latter gets determined in terms of $\alpha_0$. Solving the recursion relation for $n=1$ gives us the first-level degeneracy $m_1$ in terms of $\alpha_0$. The resulting quadratic equation for $\alpha_0$ must have a rational discriminant in order to give rational exponents. Because $m_1$ is integer, the rational discriminant must in fact be an integer, and this gives a Diophantine equation which for $\ell=0,2,3$ has only a finite, small number of solutions. Finally, by looking at the degeneracies of the first excited state in the identity character, we attempt to guess the relevant Kac-Moody algebras for these characters.

We have not yet been able to explain the close correspondence between $\ell=0$ and $\ell=2$ two-character theories, which includes a relation between central charges and conformal dimensions, and this deserves further investigation. It would also be interesting to settle the question of existence of two-character theories with $\ell>2$, other than tensor-product theories. A few sporadic cases were listed in Ref.\cite{Naculich:1988xv}. Similar questions may be addressed for three-character theories. 

By itself, the approach of classifying RCFT through modular-invariant differential equations, first proposed in Ref.\cite{Mathur:1988na}, becomes prohibitively difficult for large numbers of characters. However in recent times there have been new insights into the origin of these equations due to the seminal work of Y. Zhu \cite{Zhu:1996} that directly relates them to null vectors for chiral algebras. A hybrid of the differential-equation and null-vector approaches should be useful in taking further the classification programme. Also one may try to relate it to the general theory of vector-valued modular forms, which has been explored in recent years \cite{Bantay:2007zz,Gannon:2013jua}. Finally, it would be interesting to link this approach to a number of recent general observations about the spectrum of RCFT (for example, in Refs.\cite{Hellerman:2009bu, Keller:2014xba}) as well as to AdS$_3$ gravity on the lines of Ref.\cite{Castro:2011zq}.

\section*{Appendix A: Degeneracies for selected $\ell=2$ examples}

In this Appendix we present computational details for two selected examples out of the list in \eref{ell.2.list}. In each case we exhibit the quantities $m^{(i)}_n=\frac{a_n^{(i)}}{a_0^{(i)}}$ resulting from solving the differential equation for both the identity and non-identity character (corresponding to the superscript $(0)$ or $(1)$ respectively). We have calculated these to very high levels ($n=5000$ in the potentially interesting cases) but the resulting numbers are too large to display. Therefore we selected the range $n=31$ to 40 for the table.

The first case corresponds to $\tN=96$ and a central charge $c=\frac{118}{5}$. This case belongs in the list \eref{ell.2.list} because the first-level degeneracy above the identity is a positive integer, $m_1=59$. On consideration of higher-level degeneracies it survives into the list \eref{nearfinal} and makes it to our final list of potential CFT characters. The $m_n$ values in a selected range are presented in Table \ref{table-candidate}. Note that the column $m_n^{(0)}$ contains only positive integers, though there is no theoretical reason why a solution of the corresponding differential equation should have this property. Had just one of these numbers taken a negative integer or fractional value, the theory would have been eliminated from consideration. The column $m_n^{(1)}$ displays the ratios of excited- to ground-state degeneracies for the non-identity character. We see that the denominators are consistently made up of divisors of $551=29\times 19$. Most of the time it is 551 that occurs, but in one of the entries we find a smaller denominator of 29. An additional factor of 59 in the denominators occurs already at low levels (below those displayed in the table). Therefore the minimum degeneracy of the non-trivial primary must be $19\times 29\times 59=32509$, as displayed in Table 1. Scanning the entire list of denominators up to very high levels, we find that no denominator higher than 32509 occurs and most of them are in fact equal to 551, with sporadic occurrences of 29 or 19. It must be admitted that looking at examples up to level 5000 or any other arbitrarily chosen number cannot really rule out additional denominators. What seems more significant is the stability of these denominators over a significant range of levels, in contrast to the badly-behaved case that we discuss next which is generic in most of the theories we have ruled out.

\begin{table}[ht]
\centering
\begin{tabular}{|c||c|c|}
\hline
Level $n$ & $m_n^{(0)}$ & $m_n^{(1)}$ \\
\hline
31 &  20752043793985383874961606068 & $\frac{1564542557330629068281531726}{551}$\\
\hline
32 & 62556562614055268468241703534 & $\frac{4574584512950683479429895788}{551}$\\
\hline
33 & 185347559161586133919429897918 &  $\frac{13164911860456917137265957048}{551}$\\
\hline
34 & 540189947233095631008373977634 &  $\frac{37315204601270073786174058188}{551}$\\ 
\hline
35 &  1549779301897614070570248091958 & $\frac{104239828612481860968332126425}{551}$\\
\hline
36 &  4379777930153124757584558630242 &  $\frac{287157963869233828764755209483}{551}$ \\
\hline
37 & 12200219586436272491903360820134 &  $\frac{780529595773022969463611623710}{551}$\\
\hline
38 & 33517549085536688506182248079224 &  $\frac{2094428343912644282700433958826}{551}$\\
\hline
39 &  90866680495320339363039740822274 & $\frac{5550872031419013642027295371750}{551}$ \\
\hline
40 &243213786542971351797987467981012 &  $\frac{765106276556352177608129536210}{29}$\\
\hline
\end{tabular}
\caption{Degeneracies at levels 31 to 40 of the identity and non-identity characters for a candidate theory with $c=\frac{118}{5}$.}
\label{table-candidate}
\end{table}

This is the case of $\tN=126$, corresponding to $c=28$. This value belongs in the list \eref{ell.2.list} which means the first-level degeneracy above the identity is a positive integer (in this case, $m_1^{(0)}=4$). However the case fails to make it to the ``short-list'' \eref{nearfinal} because the higher degeneracies do not fulfil the same criterion. This is evident in Table \ref{table-noncandidate}. We see that the degeneracies of the identity character are all negativenon-negative integers. At the same time the degeneracies above the non-identity primary are rational numbers with higher and higher denominators, in sharp contrast to the ``stabilisation'' seen in the Table \ref{table-candidate}. Thus we are unable to assign a definite degeneracy to the ground state of that character such that the higher degeneracies become integers.
 
\begin{table}[ht]
\centering
\begin{tabular}{|c||c|c|}
\hline
$n$ & $m_n^{(0)}$ & $m_n^{(1)}$ \\
\hline
 31 & \tiny $ -9020070751415474425300591376720$ & \tiny$\frac{12252689835043232257816487824200227362868720235395449849336793432064}{57088849110945344189764550802020501794795} $\\
\hline
32 & \tiny$-30177130382168494210526497554425$ & \tiny$\frac{3029141734890435262819977323331703800487124697087256711248206453328}{4391449931611180322289580830924653984215}$\\
\hline
33 & \tiny$-99055059687940443115457872020232$ &   \tiny$\frac{2386194864901205563895970191499911446631818526240479685142013459298832}{1095067923855406147640029110838756898063795} $\\
\hline
34 &  \tiny$-319291773432200536074047309757218$ &  \tiny $\frac{2330391171276712120716188121216573188369888908250464434642888021277104}{344164204640270503544009149120752167962907} $\\ 
\hline
35 & \tiny$-1011496066705677909482616733781048$ & \tiny $\frac{55634705774736771927490446792375005047732522136300871212151701405573483712}{2686201617217311280160991408887470670950489135} $\\
\hline
36 & \tiny$-3151617444845477544907229957970635$ &  \tiny $\frac{590524503926359745811441432610963459869412056215299875421285164536488225786}{9463694928350219740874877425157396671502492491}$ \\
\hline
37 & \tiny$-9664957355170045458299851475537480$ & \tiny  $\frac{113974778388389997124130735211107679750737357960980287839991925545320482129632}{615140170342764283156867032635230783647662011915} $\\
\hline
38 & \tiny$-29190865906315262343516268927266728$ &  \tiny $\frac{6186884360826328831346717071043443862817535317764560986482438859528461383960632}{11403752388662014787754227297314662989160503451655}$\\
\hline
39 & \tiny$-86883862154976508910885146526418376$ & \tiny $\frac{46472511251834208044607148282163061146751006302193172347657803038195907990572032}{29649756210521238448160990973018123771817308974303} $ \\
\hline
40 & \tiny$-254990309881125092677764315638948269$ &  \tiny $\frac{662663800419673752014565625752167883586041429269686859063986904900000241396630846}{148248781052606192240804954865090618859086544871515} $\\
\hline
\end{tabular}
\caption{``Degeneracies'' at levels 31 to 40 of the identity and non-identity characters for a (failed) candidate with $c=28$.}
\label{table-noncandidate}
\end{table}

\section*{Appendix B: Dual Coxeter numbers}

For the reader's convenience, in Table \ref{table-coxeter} we have provided the dimensions, dual Coxeter numbers and central charges at level $k=1,2$ of all the simple Lie algebras.

\begin{table}[ht]
\centering
\begin{tabular}{|c|c|c|c|c|}
\hline
Lie Algebra & dim$(G)$ & $c_v$ & $c$ at $k=1$ & $c$ at $k=2$ \\
\hline
$A_n$ & $n(n+2)$ & $n+1$ & $n$ & $\frac{2n(n+2)}{n+3}$\\
\hline
$B_n$ & $n(2n+1)$ & $2n-1$ & $n+\half$ &  $2n$ \\
\hline
$C_n$ & $n(2n+1)$ & $n+1$ & $\frac{n(2n+1)}{n+2}$ & $\frac{2n(2n+1)}{n+3}$\\
\hline
$D_n$ & $n(2n-1)$ & $2n-2$ & $n$ & $2n-1$ \\
\hline
$E_6$ & 78 & 12 & 6 & $\frac{78}{7}$ \\
\hline
$E_7$ & 133 & 18 & 7 & $\frac{103}{10}$\\
\hline
$E_8$ & 248 & 30 & 8 & $\frac{31}{2}$\\
\hline
$F_4$ & 52 & 9 & $\frac{26}{5}$ & $\frac{104}{11}$\\
\hline
$G_2$ & 14 & 4 & $\frac{14}{5}$ & $\frac{14}{3}$\\
\hline
\end{tabular}
\caption{Dimensions, dual Coxeter numbers and central charges at levels $k=1,2$.}
\label{table-coxeter}
\end{table}

\section*{Conventions and useful formulae}

We have defined the Eisenstein series in the body of the paper in Eqs.(\ref{eisen}) and (\ref{eisen.2}). They are all normalised so that the first term is unity. Some useful special cases have the following series expressions:
\bea
E_2 &=& 1 -24\sum_{n=1}^\infty \frac{n q^n}{1-q^n}=1-24\sum_{n=1}^\infty \sigma_1(n)q^n\nn\\
E_4 &=& 1 +240\sum_{n=1}^\infty \frac{n^3 q^n}{1-q^n}=1+240\sum_{n=1}^\infty \sigma_3(n)q^n\nn\\
E_6 &=& 1 -504\sum_{n=1}^\infty \frac{n^5 q^n}{1-q^n}=1-504\sum_{n=1}^\infty \sigma_5(n)q^n
\eea
where
\be
\sigma_p(n)=\sum_{d|n}d^p
\ee
$E_4$ and $E_6$ can be expressed in terms of Jacobi $\theta$-functions:
\bea
E_4&=&\half \sum_{\nu=2}^4 \big(\theta_\nu(0|\tau)\big)^8\nn\\
E_6&=& \sqrt{E_4^3-\sfrac{27}{4}(\theta_2\theta_3\theta_4)^8}
\eea
Finally, the explicit expansion of these series  to a few finite orders are:
\be
\begin{split}
E_2 &= 1-24q -72 q^2 - 96 q^3 - 168 q^4\\
E_4 &= 1+240 q+2160 q^2+6720 q^3+17520 q^4\\
E_6 &= 1- 504q-16632 q^2 - 122976 q^3-532728 q^4
\end{split}
\ee

\section*{Acknowledgements}

We would like to thank Deepak Dhar, Matthias Gaberdiel, Sameer Kulkarni, Nori Iizuka and Roberto Volpato for useful discussions, and Antun Milas and Elias Kiritsis for correspondence. The work of HRH is supported by an INSPIRE Scholarship, DST, Government of India, and that of SM by a J.C. Bose Fellowship, DST, Government of India. 
We thank the people of India for their generous support for the basic sciences.


\bibliographystyle{JHEP}
\bibliography{twochar}

\end{document}